\documentclass[conference]{IEEEtran}
\IEEEoverridecommandlockouts
\usepackage{cite}
\usepackage{amsmath,amssymb,amsfonts}
\usepackage{algorithmic}
\usepackage{graphicx}
\usepackage{textcomp}
\usepackage{xcolor}
\usepackage{subfig}
\usepackage{multirow}
\usepackage{stfloats}

\def\BibTeX{{\rm B\kern-.05em{\sc i\kern-.025em b}\kern-.08em
    T\kern-.1667em\lower.7ex\hbox{E}\kern-.125emX}}
\begin{document}

\title{Modeling the Impacts of Swipe Delay on User Quality of Experience in Short Video Streaming}

\author{Duc V. Nguyen, Huyen T. T. Tran}

\maketitle

\begin{abstract}
Short video streaming platforms have gained immense popularity in recent years, transforming the way users consume video content. A critical aspect of user interaction with these platforms is the swipe gesture, which allows users to navigate through videos seamlessly. However, the delay between a user's swipe action and the subsequent video playback can significantly impact the overall user experience. This paper presents the first systematic study investigating the effects of swipe delay on user Quality of Experience (QoE) in short video streaming. In particular, we conduct a subjective quality assessment containing 132 swipe delay patterns. The obtained results show that user experience is affected not only by the swipe delay duration, but also by the number of delays and their temporal positions. A single delay of eight seconds or longer is likely to lead to user dissatisfaction. Moreover, early-session delays are less harmful to user QoE than late-session delays. Based on the findings, we propose a novel QoE model that accurately predicts user experience based on swipe delay characteristics. The proposed model demonstrates high correlation with subjective ratings, outperforming existing models in short video streaming. The dataset and code are available at $<$blinded for review$>$.
\end{abstract}


\section{Introduction}
\label{sec:intro}
Short video streaming platforms, such as TikTok, Instagram Reels, and YouTube Shorts, have revolutionized the way users consume video content. User experience with short videos, however, is different from that of traditional long-form videos in many aspects. First, short videos are typically less than 60 seconds in length and are often presented in a vertical format. Second, users interact with content more dynamically, frequently swiping to watch many videos in a session~\cite{Zhang2022}. Third, they do not explicitly choose which videos to view. Instead, videos, which are automatically played, are recommended by the platform's algorithm based on users' preferences and viewing history~\cite{Tiktok_recommend_alg}. Thus, understanding the key factors affecting users' QoE is crucial for optimizing short video streaming systems and ensuring user satisfaction.

Existing studies on the QoE of video streaming have mainly focused on scenarios where users watch a single video for a certain period. Key factors, such as initial delay and stalling, have been well defined, and their impacts have been thoroughly investigated and quantified~\cite{trans,Hossfelds,CQM}. Initial delay refers to the time between the user pressing play and the start of video playback, while stalling refers to interruptions during playback when the video freezes to buffer~\cite{QoE2014survey}. Numerous QoE models have been developed, including the standardized ITU-T P.1203 model~\cite{p1203_1}, which have demonstrated strong performance.

However, there is a critical gap in understanding user experience in the context of short video streaming. Especially, short video sessions involve frequent user interactions (swiping) that are not present in traditional video streaming. When users swipe to watch the next video, a delay between user interaction and the playback of the next video might occur due to various reasons, such as low network bandwidth, preloading strategies, or application performance issues~\cite{Phong2023}. We refer to this as \textbf{\textit{swipe delay}}. Different from initial delay and stalling, swipe delay is unique to short videos as it is associated with user interactions. Swipe delay can vary in duration and may occur multiple times within a short video session, potentially degrading user experience. Thus, it is important to understand the effects of swipe delay on users' QoE in short video streaming.



To address this gap, we conduct an extensive subjective quality assessment to evaluate the impact of swipe delay on users' QoE in short video streaming. The experiment results in a subjective dataset containing user ratings of 132 swipe delay patterns. Result analysis reveals that users' QoE is strongly affected by the delay duration, with users beginning to show dissatisfaction when the total delay exceeds 8 seconds. Given the same total delay, the user rating is reduced by approximately 0.5 Mean Opinion Score (MOS) when the number of delay increase from one to four. It is also found that delays near the end of a session cause more impact than those that occur earlier. Based on the constructed subjective dataset, we develop a novel QoE model that accurately predicts user experience based on the characteristics of swipe delay. The proposed model has high correlation with subjective ratings and outperforms models originally designed for traditional video streaming. The dataset and the proposed QoE model are made publicly available to facilitate further research in this direction. To the best of our knowledge, this is the first work on QoE assessment in short video streaming.

The rest of the paper is organized as follows. Section \ref{sec:related_work} reviews related work on short video streaming and QoE modeling. Section \ref{sec:subjective_assessment} describes the subjective experiment. Section \ref{sec:experimental_results} analyzes the impacts of key factors on the users' QoE. Section \ref{sec:proposed_model} describes the proposed QoE prediction model. Finally, Section \ref{sec:conclusion} concludes the paper.

\section{Related Work}
\label{sec:related_work}
Recently, short video streaming has received great attention from both academia and industry. Empirical studies have been conducted to understand the characteristics of commercial short video streaming platforms and user behaviors~\cite{Zhang2022,SwipeAlong, Measurement_Study_2025}. It is shown that videos are generally extremely short, with a median video length of 22 seconds, and approximately 77\% of short video sessions last less than 40 seconds. Moreover, users exhibit limited attention span, as nearly 70\% of videos are not watched to completion, leading to substantial data wastage~\cite{Zhang2022,Measurement_Study_2025}. In~\cite{SwipeAlong}, it is found that popular short video streaming platforms employ simple protocols. Typically, each video is fully downloaded before playback, and a fixed number of subsequent videos are preloaded to mitigate swipe delay.

A key challenge in short video streaming is how to minimize data wastage without affecting users' QoE. Both heuristic-based (e.g.,~\cite{nguyen2022,Phong2023,dashlet}) and learning-based solutions (e.g.,~\cite{LiveClip,Gamora,DeLoad}) have been developed.  In heuristic solutions, the optimal preloading size and bitrate of individual videos are determined based on the user's past viewing behavior and network conditions. In a learning-based approach, users' past viewing behaviors, network conditions, and video metadata are used to learn an optimal preloading policy using reinforcement learning~(e.g., ~\cite{Gamora}). To measure the users' QoE of a streaming session, previous works employ  QoE models that are originally developed for long-form video streaming. Thus, factors specific to short videos, such as swipe delay, are not considered. Also, there is no previous work that aims to assess users' QoE in the context of short video streaming.

QoE assessment in video streaming has been studied extensively in the literature~\cite{QoE2014survey}. Prior work has primarily focused on traditional streaming scenarios, in which users watch a single video per session, typically lasting from 8 seconds to 5 minutes~\cite{largescale,ICME_RMSE3}. Key factors affecting users' QoE include initial delay, stalling, and quality variation~\cite{Hossfelds,yins}. Among them, stalling has been shown to severely degrade QoE, whereas initial delay and quality variation have relatively smaller impacts~\cite{trans}. In addition, a recency effect has been observed for stalling and quality variation, whereby more recent events exert a greater impact. Many QoE models for single-video streaming have been developed~\cite{CQM,CAT}, including the standardized ITU-T P.1203 model~\cite{p1203_1}. These models are based on either analytical functions (e.g., linear~\cite{trans} and exponential~\cite{Hossfelds}) or learning-based approaches (e.g., LSTM~\cite{LSTMTran} and decision trees~\cite{p1203_1}). They have been shown to perform well on large-scale QoE datasets~\cite{QoEdatabase1,QoEdatabase2,largescale}. However, as demonstrated in Section~\ref{sec:proposed_model}, existing QoE models exhibit suboptimal performance in short video streaming sessions involving swipe delay. 

QoE of multimedia applications that involve frequent user interactions, such as web browsing, cloud gaming, and IPTV have been investigated. Previous works show that response times are one of the most important QoE factor~\cite{WebQoESurvey,CloudGameQoE}. Pioneer work~\cite{threedelaylimit} defines three critical limits of response times that affect user experience. However, it has been shown that users' tolerable threshold varies across applications. In cloud gaming, the response time should not be lower than 200 milliseconds to maintain acceptable QoE for medium and fast games~\cite{CloudGameQoE}. Meanwhile, in IPTV applications, it is found that the zapping time should be less than 0.43 seconds to guarantee an acceptable QoE~\cite{kooij2006perceived}.

\begin{figure}[t]
    \centering
    \subfloat[Video 1]{\includegraphics[width=0.15\columnwidth]{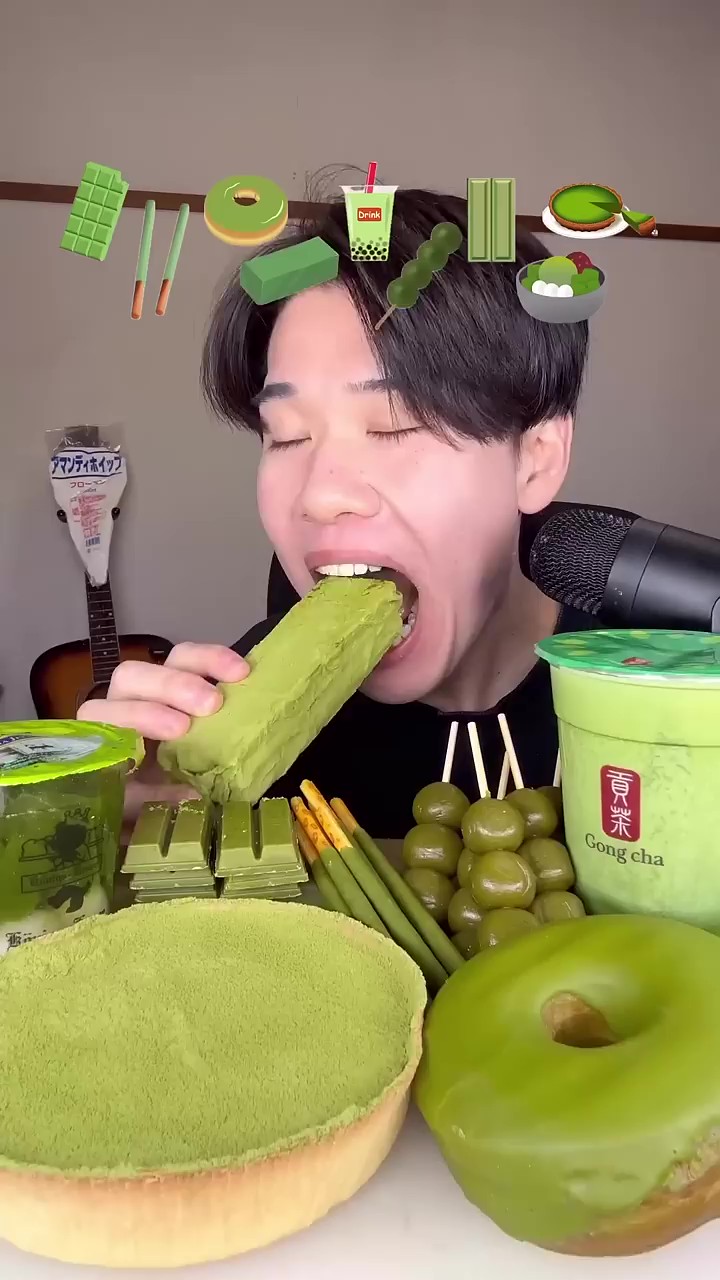}}\hfill
    \subfloat[Video 2]{\includegraphics[width=0.15\columnwidth]{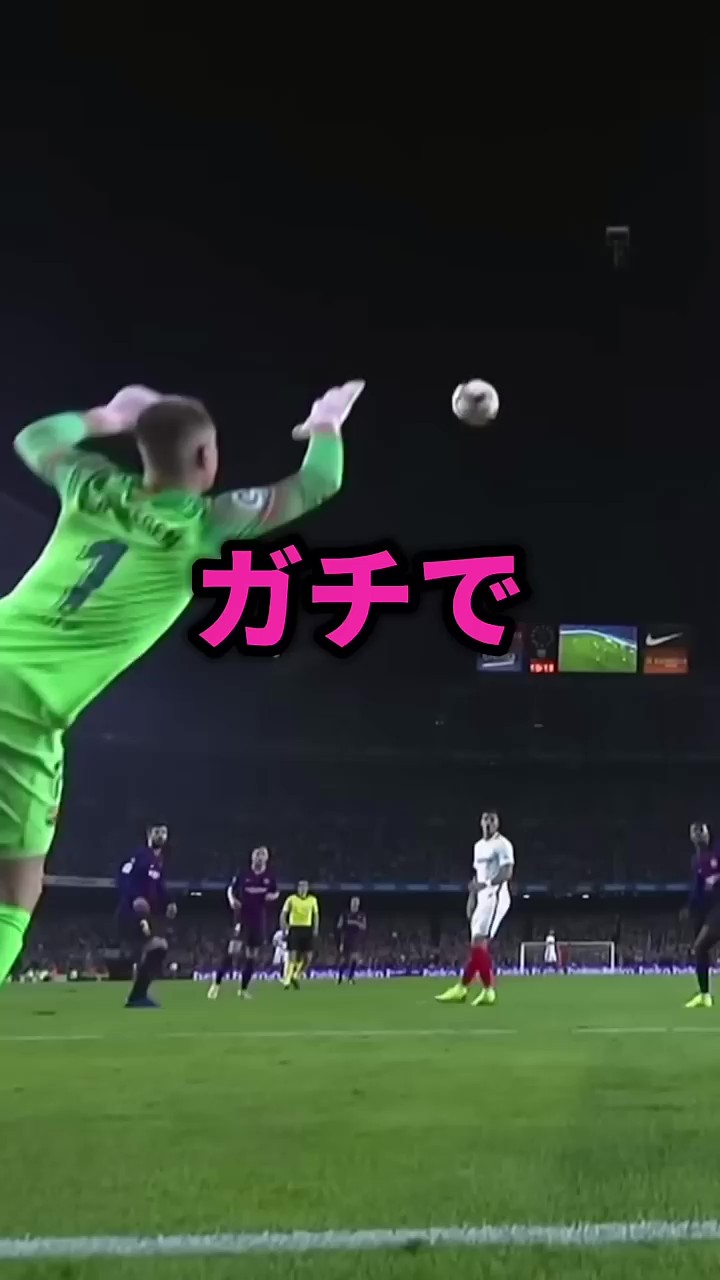}}\hfill
    \subfloat[Video 3]{\includegraphics[width=0.15\columnwidth]{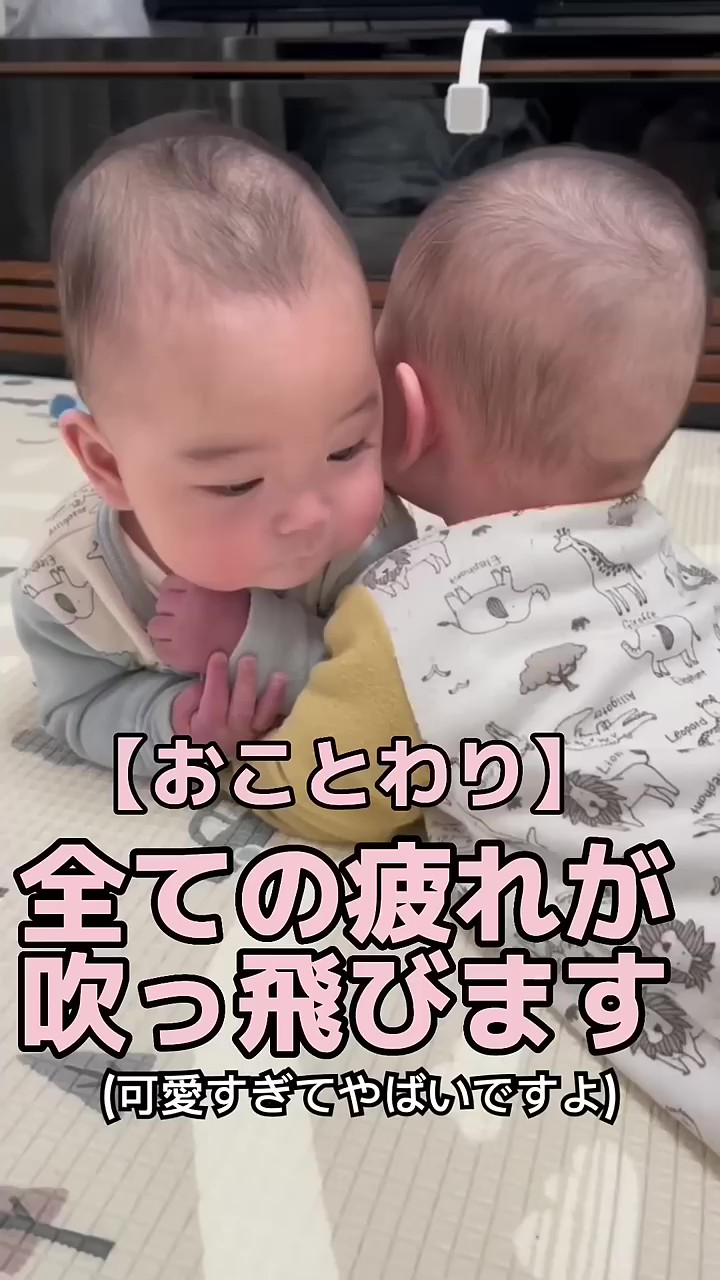}}\hfill
    \subfloat[Video 4]{\includegraphics[width=0.15\columnwidth]{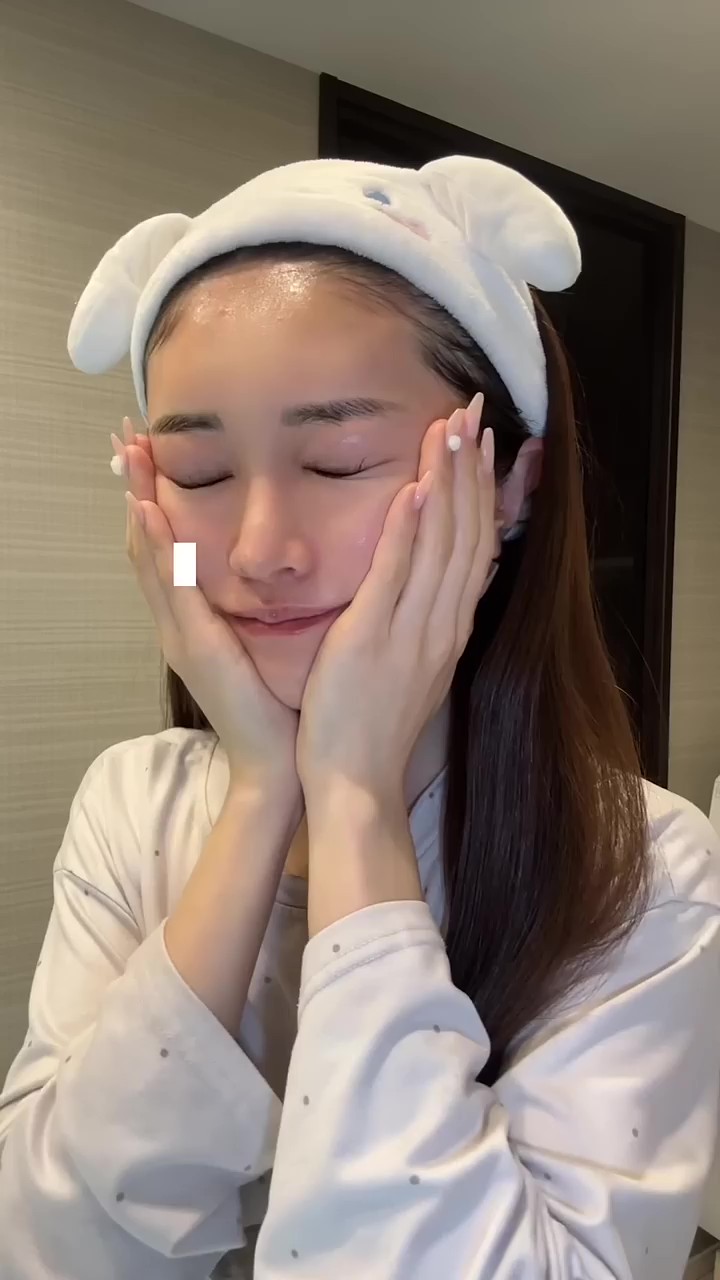}}\hfill
    \subfloat[Video 5]{\includegraphics[width=0.15\columnwidth]{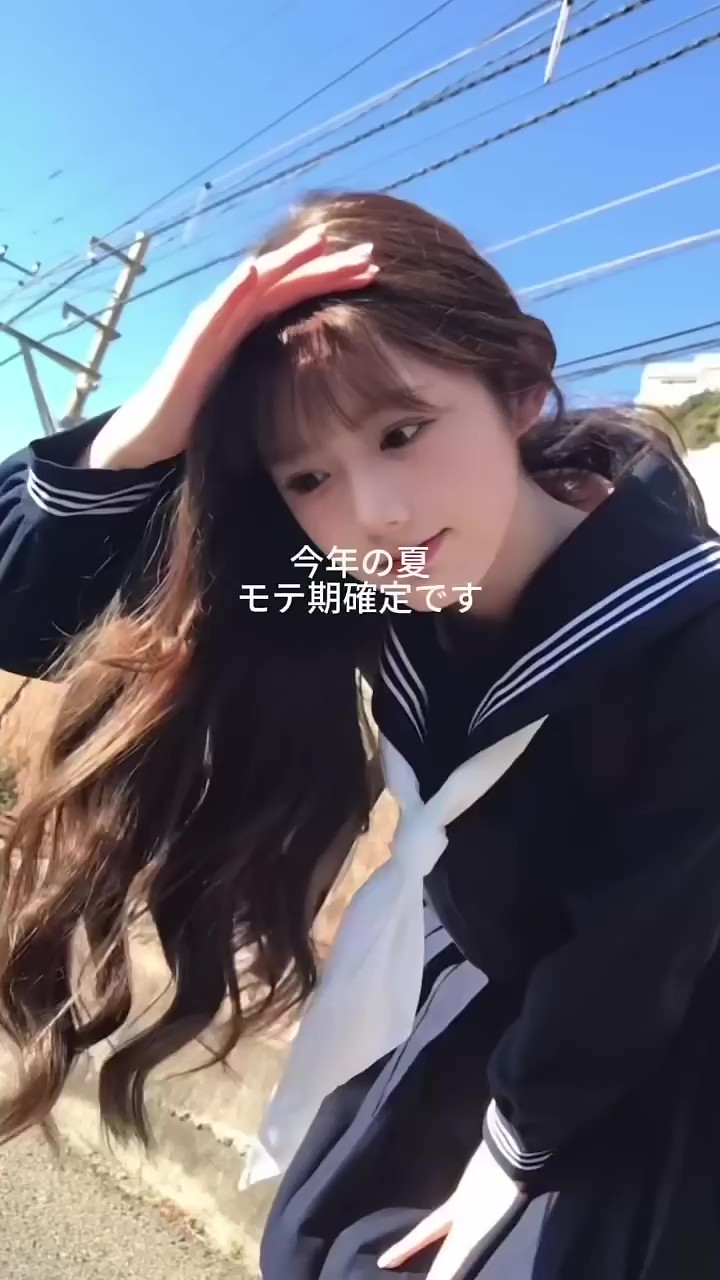}}\hfill
    \subfloat[Video 6]{\includegraphics[width=0.15\columnwidth]{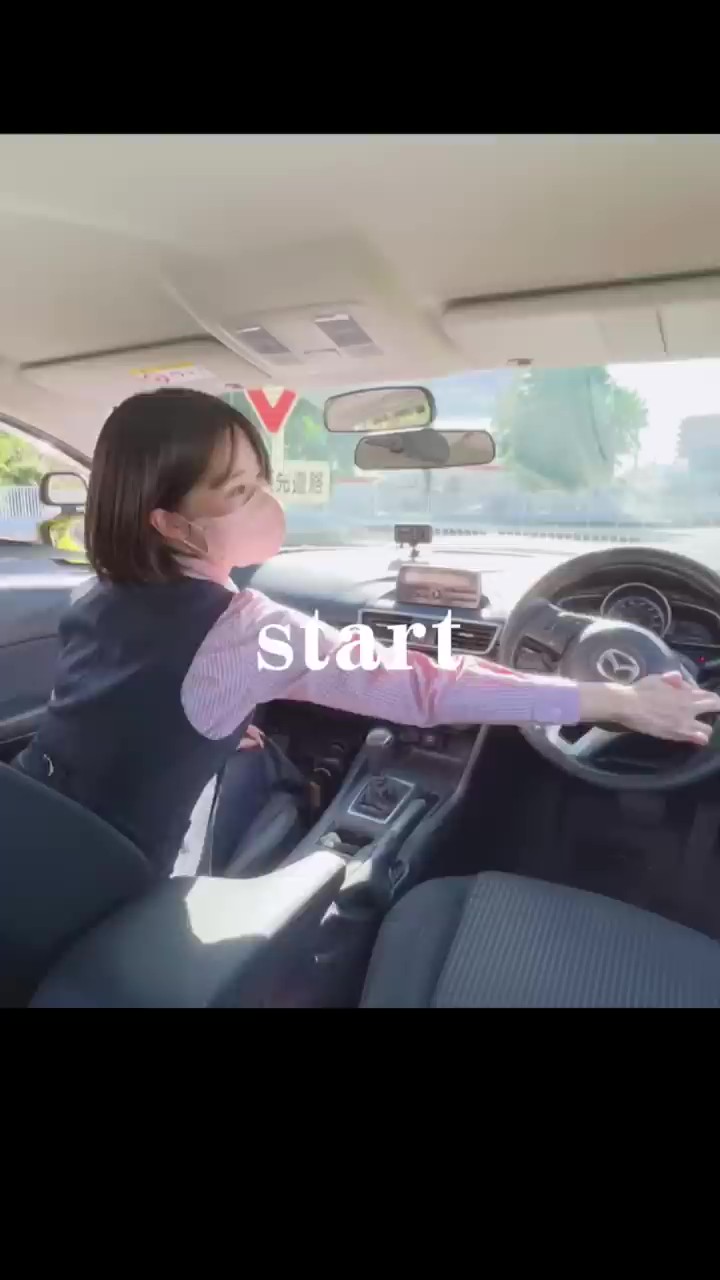}}
    \caption{Snapshots of used short videos.}
    \label{fig:video_snapshots}
\end{figure}
\begin{table}[t]
\caption{Characteristics of used short videos.}
\begin{center}
\resizebox{0.97\columnwidth}{!}{%
\begin{tabular}{|c|c|c|c|c|c|}
\hline
\multirow{2}{*}{\textbf{Video}} & \textbf{Duration} & \multirow{2}{*}{\textbf{Resolution}}  & \textbf{Bitrate} & \textbf{Frame rate} & \textbf{Content }  \\
 & \textbf{(seconds)} &  & \textbf{(kbps)} & \textbf{(fps)} & \textbf{type} \\
\hline
1 & 60 & 720$\times$1080 & 1154 & 30 & Food\\
\hline
2 & 35 & 720$\times$1080 & 1912 & 30 & Music \\
\hline
3 & 29 & 720$\times$1080 & 1086 & 30 & Baby \\
\hline
4 & 59 & 720$\times$1080 & 934 &  30 & Makeup \\
\hline
5 & 26 & 720$\times$1080 & 1802 & 30 & School \\
\hline
6 & 23 & 720$\times$1080 & 773 & 30 & Driving \\
\hline
\end{tabular}
}
\label{tab:video_characteristics}
\end{center}
\vspace{-14pt}
\end{table}

\section{Subjective Experiment}
\label{sec:subjective_assessment}
This section outlines the subjective experiment, including the design, environment, methodology, and procedure.
\subsection{Test Design}

\begin{figure}[t]
    \centering
    \includegraphics[width=\columnwidth]{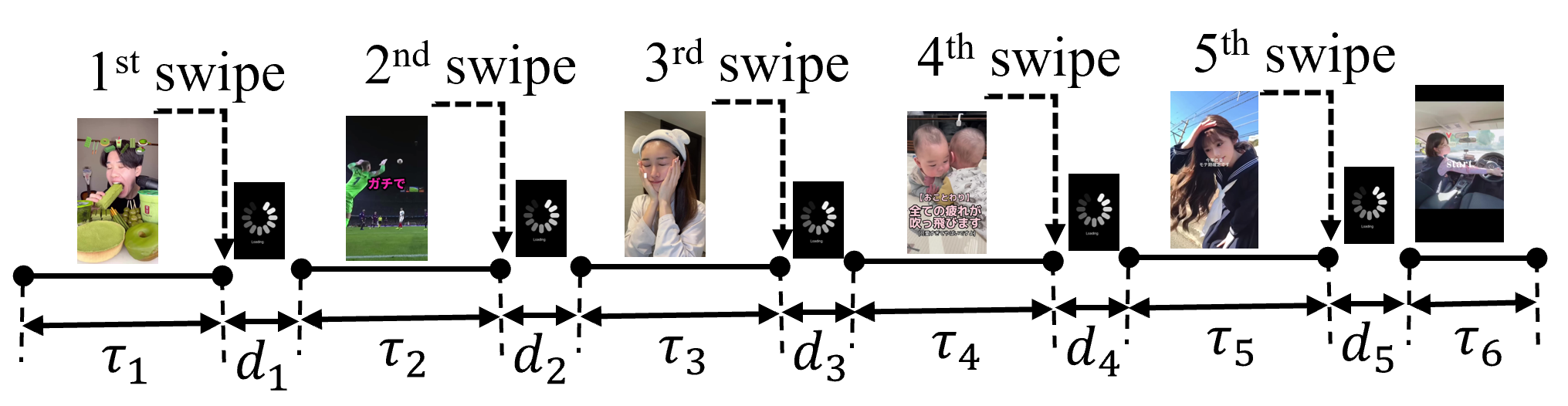}
    \caption{Illustration of a test stimulus. A loading screen is shown in case of a swipe delay.}
    \label{fig:test_session}
\end{figure}

\begin{table}[t]
\caption{Swipe delay patterns given a total swipe delay time of $D$ seconds. $d_i$ denotes the swipe delay duration after video $i~(1\leq i \leq 5)$. $N^d$ denotes the number of swipe delays.}
\begin{center}
\begin{tabular}{|c|c|c|c|c|c|c|}
\hline
\textbf{Pattern ID} & \textbf{$d_1$} & \textbf{$d_2$} & \textbf{$d_3$} & \textbf{$d_4$} & \textbf{$d_5$}  & $N^d$ \\ 
\hline
$P_1$ & $D$ & 0 & 0 & 0 & 0 & 1 \\
\hline
$P_2$ & 0 & 0 & 0 & 0 & $D$ & 1 \\
\hline
$P_3$ & \textbf{$D/2$} & \textbf{$D/2$} & 0 & 0 & 0 & 2 \\
\hline
$P_4$ & 0 & 0 & 0 & \textbf{$D/2$} & \textbf{$D/2$}  & 2\\
\hline
$P_5$ & \textbf{$D/3$} & \textbf{$D/3$} & \textbf{$D/3$} & 0 & 0 & 3 \\
\hline
$P_6$ & 0 & 0 & \textbf{$D/3$} & \textbf{$D/3$} & \textbf{$D/3$} & 3 \\
\hline
$P_7$ & \textbf{$D/4$} & \textbf{$D/4$} & \textbf{$D/4$} & \textbf{$D/4$} & 0&  4 \\
\hline
$P_8$ & 0 & \textbf{$D/4$} & \textbf{$D/4$} & \textbf{$D/4$} & \textbf{$D/4$} &  4 \\
\hline
\end{tabular}
\label{tab:delay_patterns}
\end{center}
\vspace{-14pt}
\end{table}

\begin{table}[t]
\caption{Summary of experimental conditions.}
\begin{center}
\begin{tabular}{|c|c|}
\hline
\textbf{Parameter} & \textbf{Value} \\
\hline
Number of Videos ($N$) & 6 \\
\hline
Viewing Duration in seconds ($\tau$) & 2, 4, 8, 16 \\
\hline
Total Delay Time in seconds ($D$) & 0, 1, 4, 8, 16 \\
\hline
Number of Delays ($N^d$) & 1, 2, 3, 4  \\
\hline
Delay Patterns & 8 (Table \ref{tab:delay_patterns}) \\
\hline
\end{tabular}
\label{tab:test_condition}
\end{center}
\vspace{-14pt}
\end{table}

\begin{figure}[t]
    \centering
    \subfloat[Swipe-down indicator]{\includegraphics[width=0.446\columnwidth]{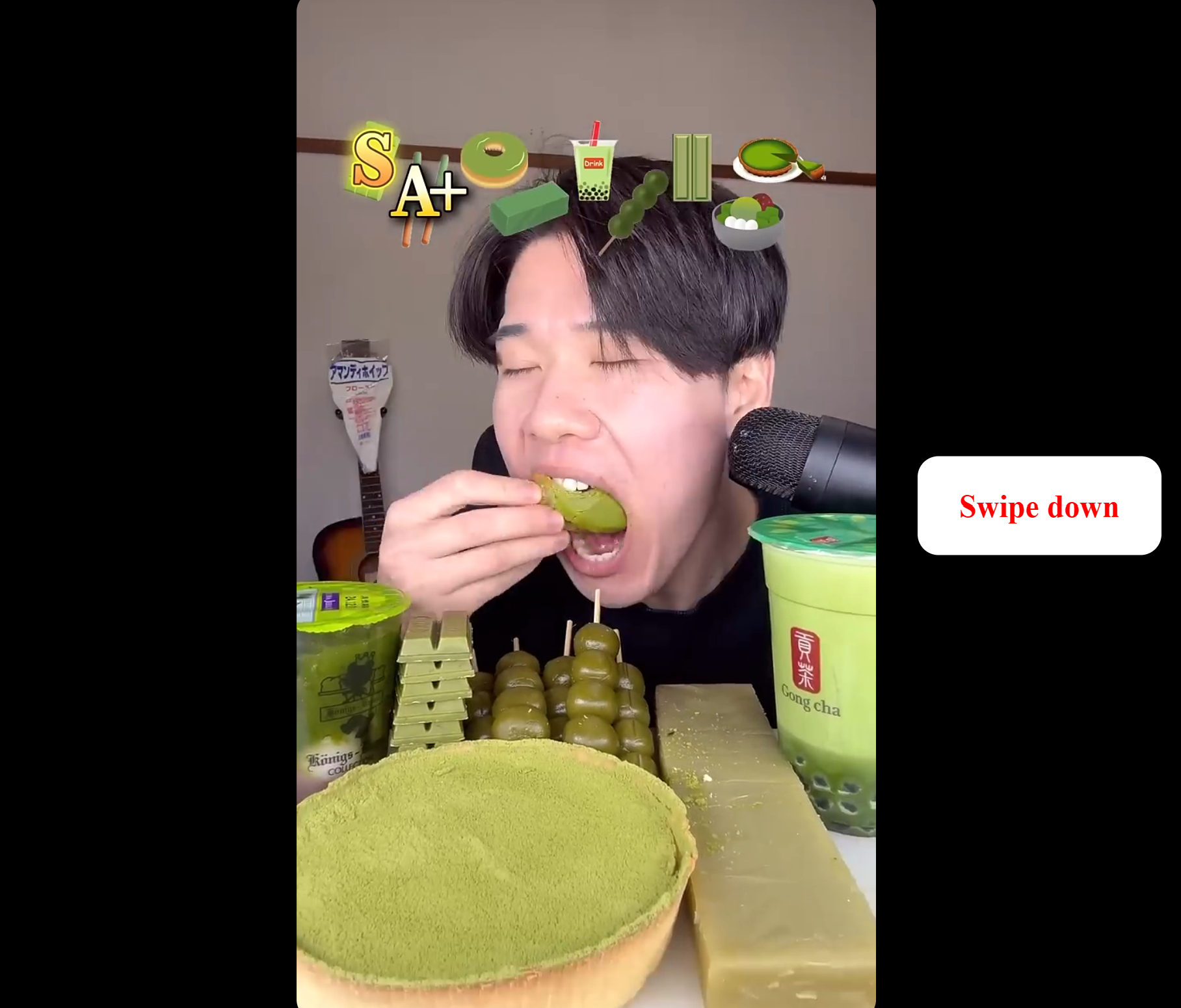}}\hfill
    \subfloat[Rating interface]{\includegraphics[width=0.47\columnwidth]{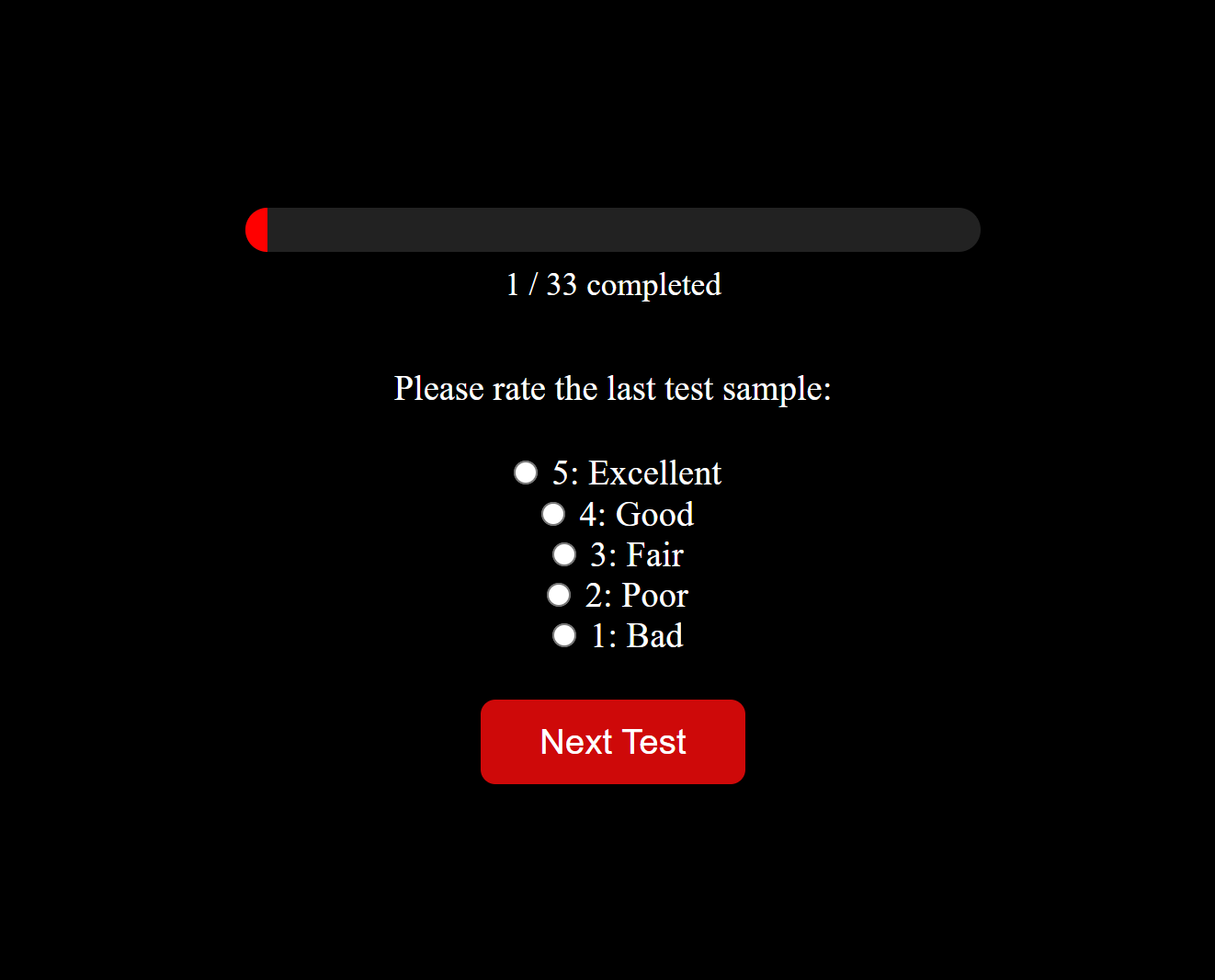}}\hfill
    \caption{Screenshots of application interface.}
    \label{fig:screenshot}
    \vspace{-14pt}
\end{figure}

In our experiment, six short videos of diverse content types are used to simulate short video streaming sessions. Snapshots of the videos are shown in Fig.~\ref{fig:video_snapshots}, and their key characteristics are summarized in Table \ref{tab:video_characteristics}. The videos have durations ranging from 23 to 60 seconds, a resolution of 720$\times$1080, and a frame rate of 30 fps.

Each stimulus is defined by the viewing duration per video and the swipe delay duration after each video, as illustrated in Fig.~\ref{fig:test_session}. When a swipe delay occurs, a loading screen is displayed during the delay period. Let $\tau_i$ and $d_i$ respectively denote the viewing duration and swipe delay of video $i \in \{1, \dots, 6\}$. To examine the impact of the time interval between delays, the viewing durations of the first five videos in each stimulus are set to be identical (i.e., $\tau_1=\tau_2=\tau_3=\tau_4=\tau_5=\tau$), ensuring a fixed inter-delay interval. Since no swipe delay occurs after the last video, the final video is always assigned a viewing duration of 1 second (i.e., $\tau_6=1$), with the corresponding delay set to zero (i.e., $d_6=0$) for all stimuli.

Given a total delay time of $D$ seconds, we designed eight delay patterns with the number of delays ranging from one to four, as summarized in Table \ref{tab:delay_patterns}. These delay patterns can also be organized into four categories based on the number  of delays:
\begin{itemize}
    \item Single-delay: Only a single delay occurs either after the first video (i.e., $P_1$) or after the fifth video (i.e., $P_2$).
    \item Double-delay: Two delays occur either after the first and second videos (i.e., $P_3$) or after the fourth and fifth videos (i.e., $P_4$)
    \item Triple-delay: Three delays occur either after the first, second, and third videos (i.e., $P_5$) or after the third, fourth, and fifth videos (i.e., $P_6$)
    \item Quadruple-delay: Four delays occur either after the first, second, third, and fourth videos (i.e., $P_7$) or after the second, third, fourth, and fifth videos (i.e., $P_8$).
\end{itemize}
This design allows us to systematically investigate the effects of not only the delay duration, the number of delays, but also the positions of the delay (e.g, at the beginning vs. end).

Table \ref{tab:test_condition} provides a summary of the experimental conditions employed in this study. The conditions are chosen to cover a wide range of swipe delay characteristics, including the video's viewing duration, total swipe delay, number of swipe delays, and delay positions. To reflect varying user engagement levels, we consider four different viewing durations of 2, 4, 8, and 16 seconds. We consider five values for the total swipe delay duration of 0, 1, 4, 8, and 16 seconds. The duration of the single swipe delay is between 0.25 seconds~($D=1$, patterns $P_7/P_8$) and 16 seconds ($D=16$, patterns $P_1/P_2$). It should be noted that in the case of no delay (i.e., $D=0$), there is only one delay pattern per viewing duration as all the patterns in Table~\ref{tab:delay_patterns} become identical. In total, our evaluation consists of 132 unique test stimuli with durations between 11 and 97 seconds and an average duration of 45.8 seconds. 

\subsection{Test Environment, Methodology, and Procedure}
\color{black}
To conduct the subjective evaluation, we developed a Web-based application that simulates a short video service interface. The application presents a predefined list of videos in a vertically scrollable layout, allowing participants to navigate through the videos using mouse or touch gestures. All videos are preloaded onto the user's device to ensure accurate simulation of the designed delay patterns. To incorporate user interaction, a swipe-down indicator is displayed outside the video area at a predefined time, prompting the user to navigate to the next video (see Fig.~\ref{fig:screenshot}a for an interface screenshot). For example, when the viewing duration per video is set to 2 seconds, the swipe-down indicator appears 2 seconds after the start of each video. 

We use the Absolute Category Rating (ACR) method to measure the subjective QoE of test stimuli. In particular, each stimulus is presented one at a time and rated independently on a 5-point scale: 5 (Excellent), 4 (Good), 3 (Fair), 2 (Poor), and 1 (Bad)~\cite{ITU-T-P910}. Fig.~\ref{fig:screenshot}b shows a screenshot of the rating interface. We recruited 40 participants aged between 20 and 35, all of whom had between 1 and 5 years of experience using short video services. The experiment is conducted in a controlled laboratory environment with consistent lighting and minimal distractions. All tests are performed on desktop computers equipped with 24-inch monitors. 

The subjective evaluation procedure consists of two phases: a training phase and a test phase. In the training phase, participants are first briefed on the test objectives, the rating scale, and the evaluation procedure. They are then asked to rate four training stimuli to familiarize themselves with the test environment. Upon completing the training phase, participants advance to the test phase and rate the stimuli, which are presented in a randomized order to mitigate potential order effects. To mitigate participant fatigue, the stimuli are divided into five subsets, each rated on a different day. Also, participants are asked to take a 3-minute break every 20 minutes of testing. Once the evaluation is completed, post-experimental screening using Pearson linear correlation is performed, and participants with a correlation smaller than a discard threshold of 0.75 are removed~\cite{ITU-T-P910}. After screening, each stimulus is rated by twenty valid participants. The MOS of a test stimulus is computed as the average rating over all valid participants.

\color{black}

\begin{figure*}[t]
    \centering
    \subfloat[]{
        \centering
        \includegraphics[width=0.33\textwidth]{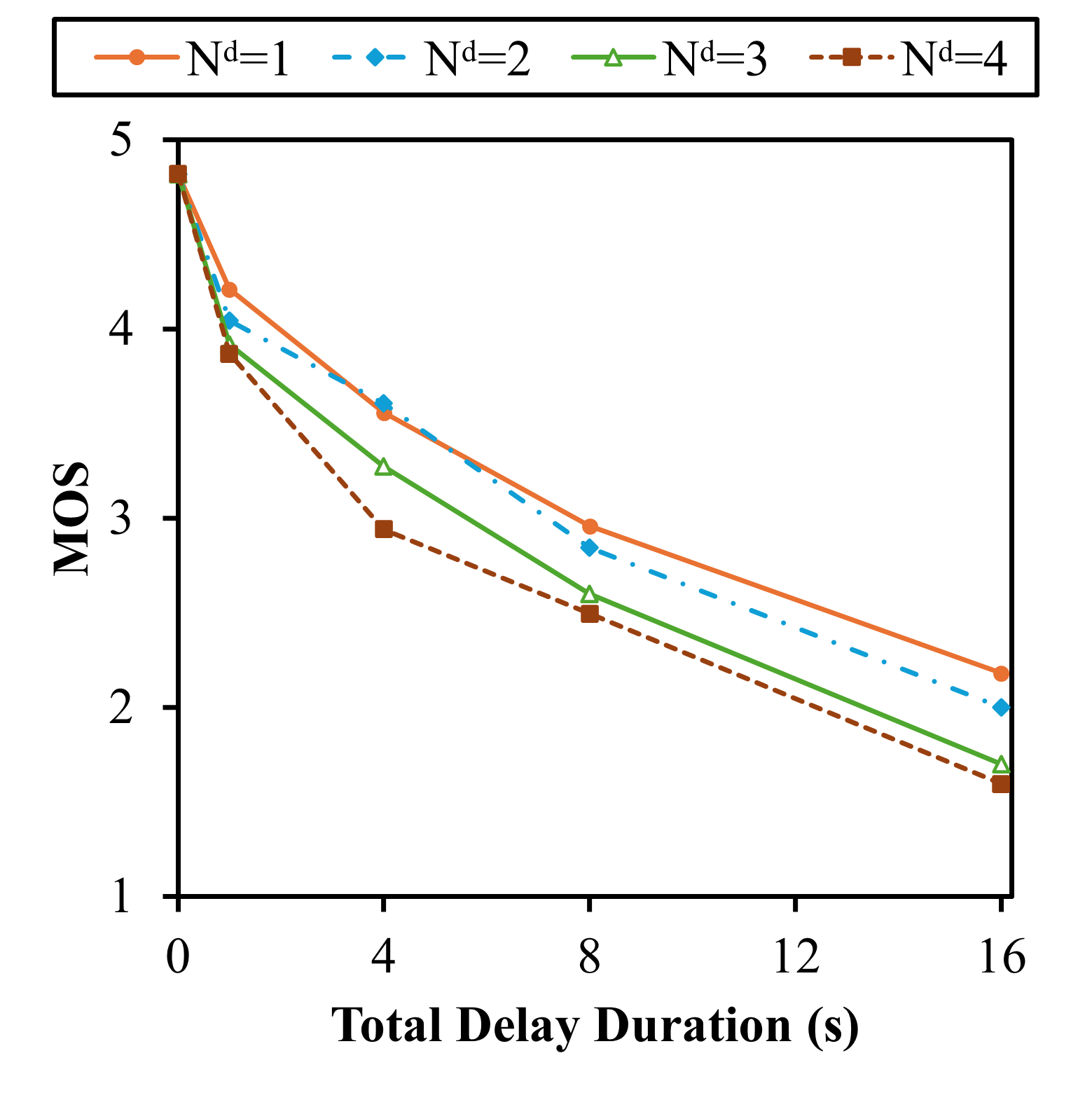}
        \label{fig:total_delay_impact}
    }
    \subfloat[]{
    \centering
    \includegraphics[width=0.33\textwidth]{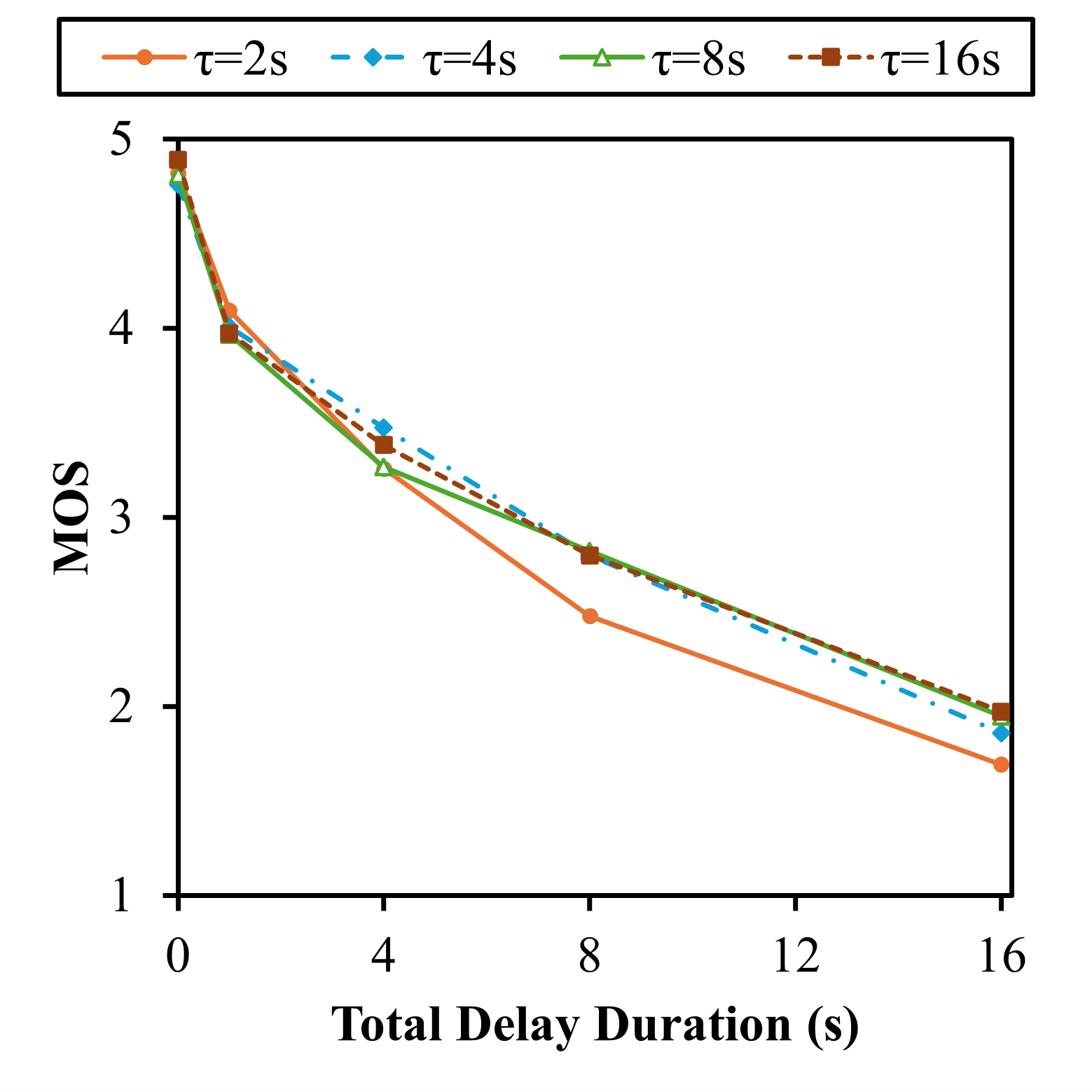}
     \label{fig:viewtime_impact}
    }
    \subfloat[]{
    \centering
    \includegraphics[width=0.33\textwidth]{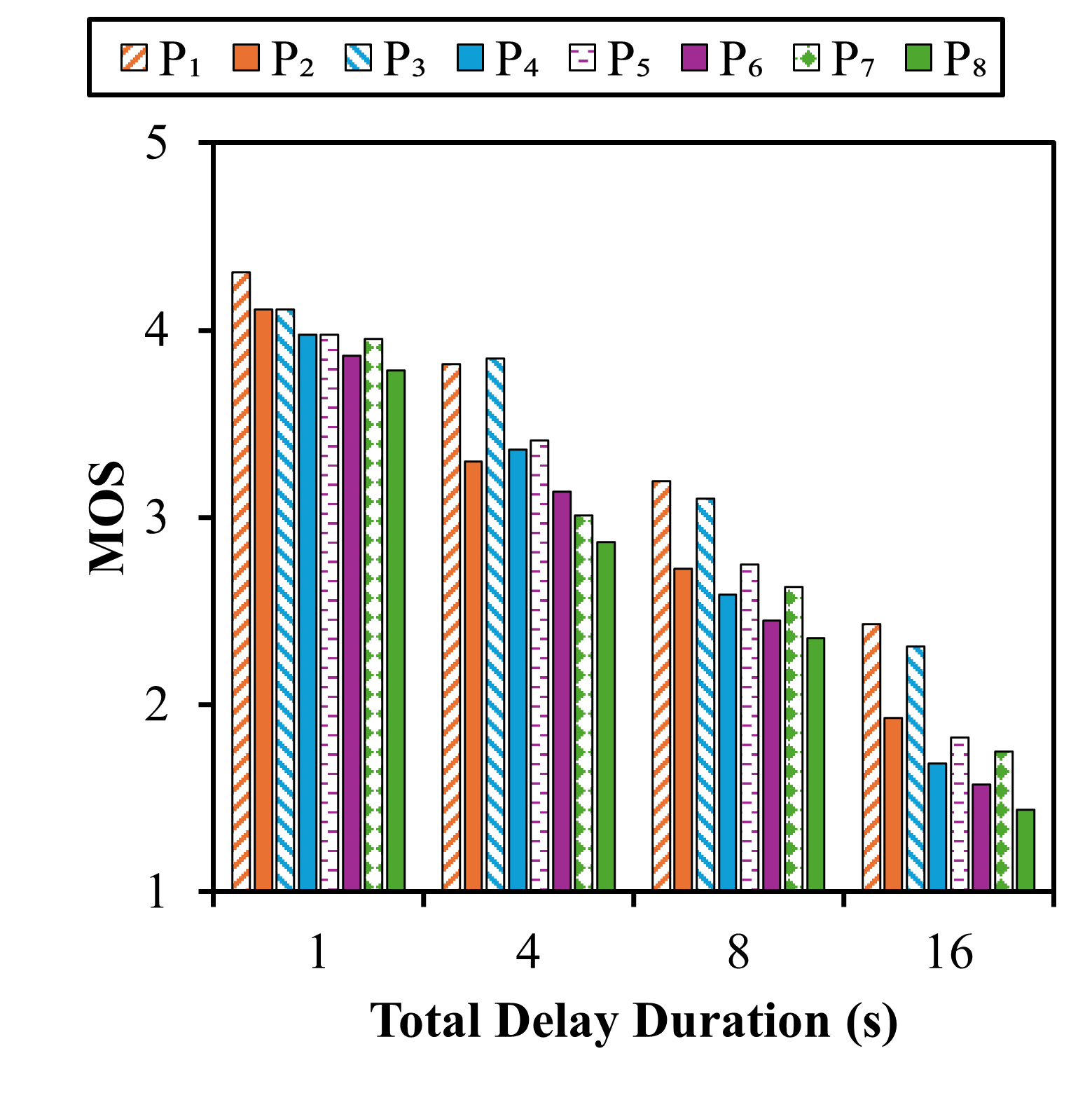}
     \label{fig:delay_position_impact}
    }
    
    \hfill
    \caption{Impacts of key characteristics: total delay duration, number of delays, time interval between delays, and delay positions on user QoE.}
    \label{fig:factor_impact}
    \vspace{-14pt}
\end{figure*}

\begin{figure}[t]
    \centering
    \subfloat[]{%
        \includegraphics[width=0.49\columnwidth]{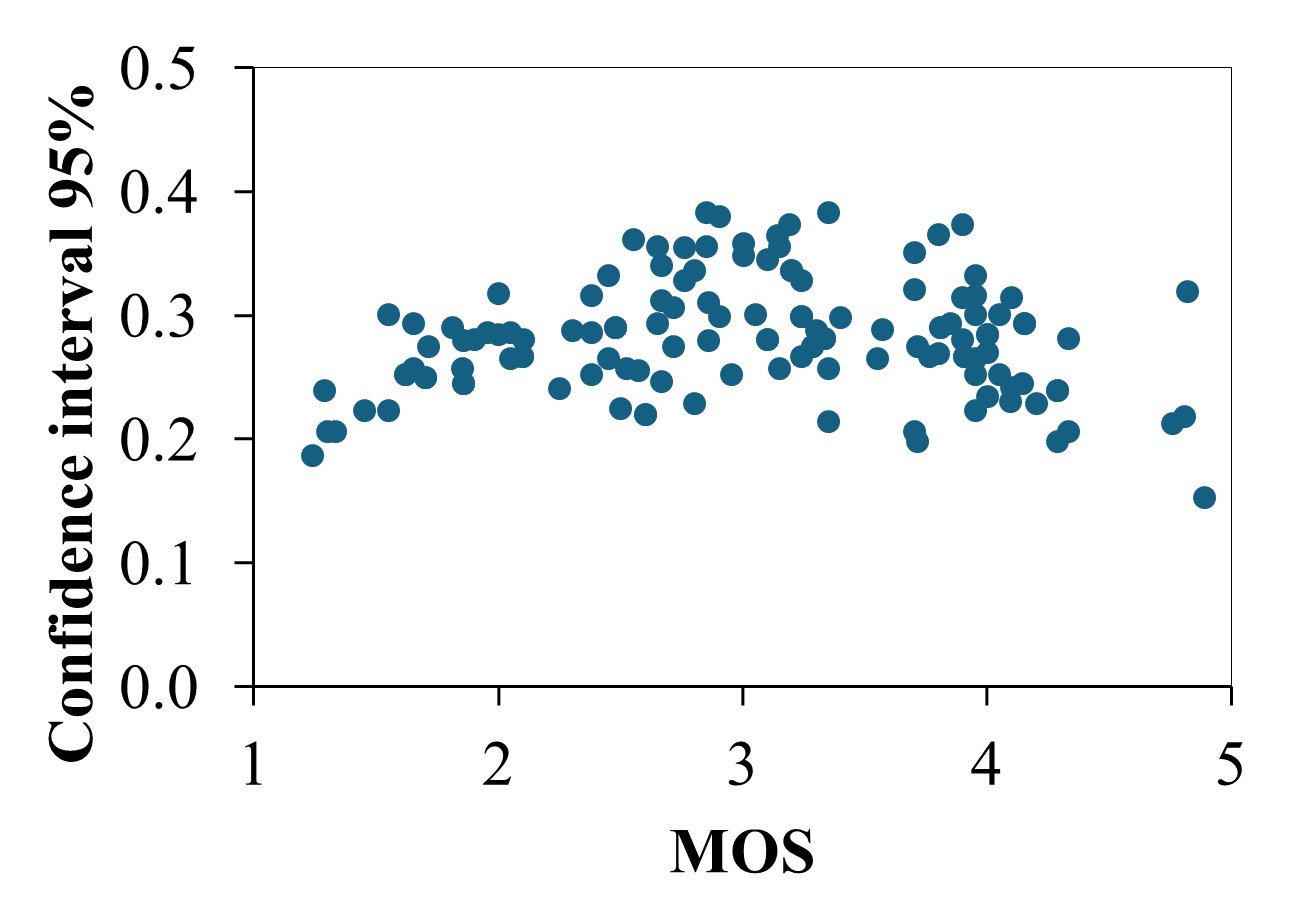}%
        \label{fig:mos_ci}%
    }
    \hfill
    \subfloat[]{%
        \includegraphics[width=0.49\columnwidth]{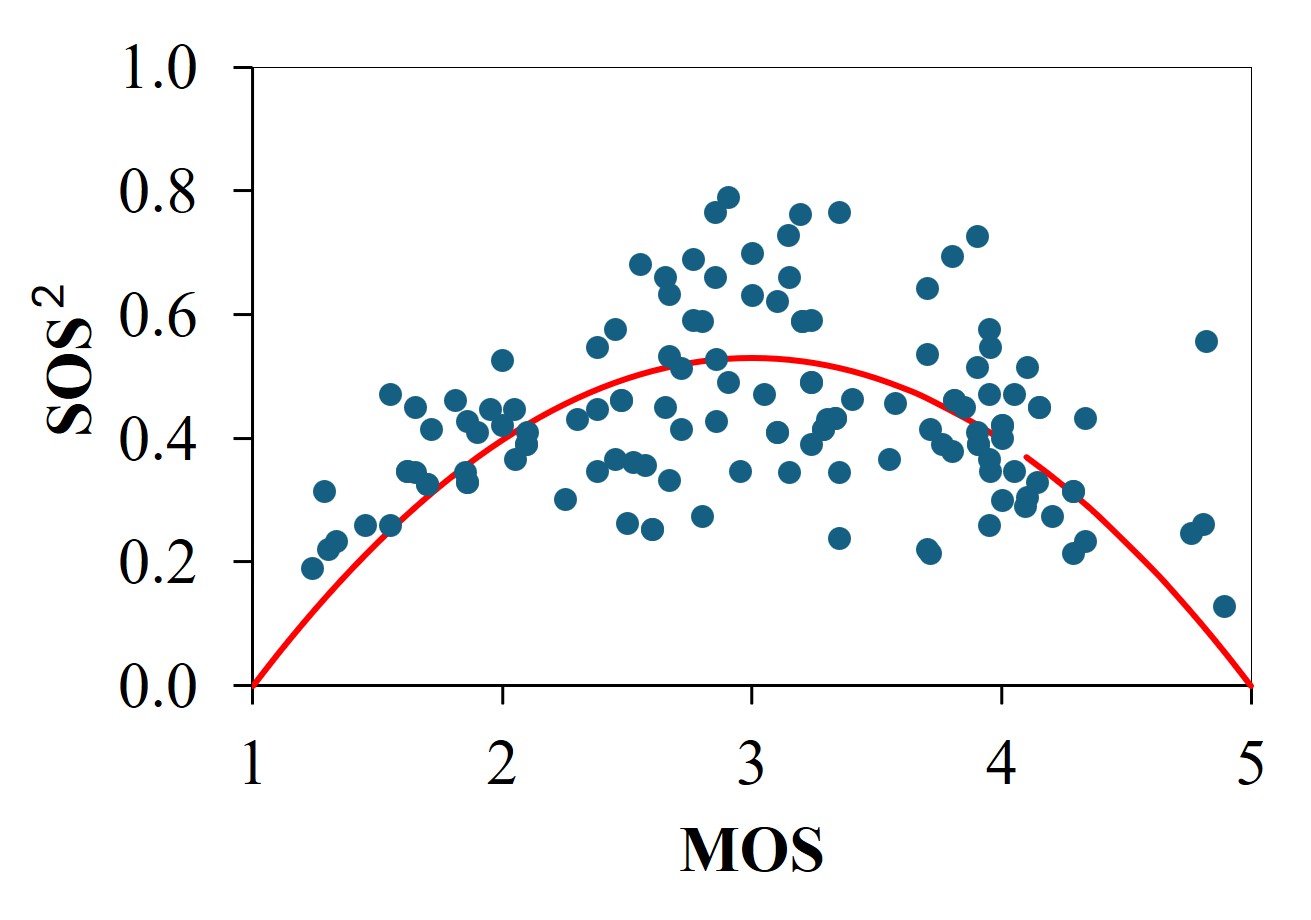}%
        \label{fig:sos}%
    }
    \caption{(a) 95\% confidence intervals of MOSs. (b) Scatter plot and fitted curve of MOS and SOS$^2$.}
    \label{fig:two_figures}
    \vspace{-14pt}
\end{figure}

\section{Results and Analysis}
\label{sec:experimental_results}

\subsection{Reliability of Subjective Ratings}
To examine the reliability of the obtained user ratings, we first analyze the 95\% confidence intervals (CIs) of the MOSs, as shown in Fig.~\ref{fig:mos_ci}. The CIs range from 0.15 to 0.38, with smaller values occurring near the extremes of the MOS scale (i.e., MOS = 1 and MOS = 5). The average CI is 0.28, which is consistent with values reported in previous subjective quality assessment studies~\cite{CI_2, ICME_RMSE3}.

We further evaluate the reliability of the MOSs by analyzing the Standard Deviation of Opinion Scores (SOS)~\cite{SOS}. The results, shown in Fig.~\ref{fig:sos}, yield an SOS parameter $a=0.132$, which falls within the range typically observed in prior work~\cite{SOS_006to018,SOS_03,SOS_0244}. This confirms the reliability of the obtained user ratings.

\subsection{Impact of Swipe Delay Characteristics on User QoE}
In this part, we analyze the effects of key characteristics, namely the total swipe delay duration ($D$), the number of swipe delays ($N^d$), the time interval between delays, and the delay positions on the users' QoE.

\subsubsection{Total delay duration and number of delays} Figure~\ref{fig:total_delay_impact} shows the average MOSs of stimuli with the same total delay duration and number of delays. It is shown that the user QoE is affected not only by the total delay duration but also by the number of swipe delays. Specifically, the MOS decreases as the total delay duration and/or the number of delays increases. When the total delay duration is 1 second, the MOS is in the range (3.9$\sim$4.2) for all the numbers of delays, indicating a good user experience. When the total delay increases to 4 seconds, the MOS is still higher than 3.0 in the cases of one to three delays, but becomes lower than 3.0 in the case of four delays. Thus, when $D=4s$, an acceptable QoE can still be achieved if there are no more than three delays. When the total delay duration exceeds 8 seconds, the MOS drops below 3.0. In particular, the MOS is in the range 2.5$\sim$2.9 at $D=8s$ and in the range 1.6$\sim$2.2 at $D=16s$. This result implies that a single delay of 8 seconds or higher generally leads to poor user experience. It is also shown that, given the same total delay duration, the MOS becomes smaller as the number of delays increases. In particular, the MOS is reduced by 0.3$\sim$0.6 as the number of delays increases from one to four.

\subsubsection{Time interval between delays} Figure \ref{fig:viewtime_impact} shows the average MOSs of stimuli with the same total delay duration and the time interval between consecutive delays (or the viewing duration per video $\tau$). It is shown that, when the total delay duration is 4 seconds or less, the MOSs are similar across all time intervals. This result indicates that, for short delays, users are relatively insensitive to the time interval between delays. For $D=\{8, 16\}$ seconds, the MOSs remain comparable for viewing durations ranging from $\tau=4$ to $\tau=16$ seconds, whereas the MOSs at $\tau=2$ seconds are significantly lower. This suggests that, for long delays, delays occurring in close temporal proximity (e.g., 2 seconds apart) lead to more severe degradation of QoE. In contrast, when delays are separated by longer intervals (e.g., $\geq$ 4 seconds), the impact of the time interval between delays becomes less profound.


\subsubsection{Delay position} Figure \ref{fig:delay_position_impact} compares the MOSs of eight swipe delay patterns as described in Table~\ref{tab:delay_patterns}. It can be observed that, given the same number of delays and total delay duration, the MOS scores are higher when the delays occur at the beginning of the session compared to when they appear toward the end of the session. Specifically, the MOS of pattern $P_1$ (single swipe delay at video \#1) is higher than that of pattern $P_2$ (single swipe delay at video \#5). Similar results can be seen when comparing patterns $P_3$ and $P_4$, $P_5$ and $P_6$, and $P_7$ and $P_8$. For the patterns $P_7$ and $P_8$ with four delay events, the trend persists even though the difference is only at the first and fifth videos. This result suggests there is a recency effect regarding the temporal position of the swipe delay, in which delays at the beginning can be forgotten as users continue to watch more videos, while delays at the end may leave a lasting negative impression. 


\section{QoE Prediction Model}\label{sec:proposed_model}

In this section, we present a novel QoE prediction model that quantifies the impact of swipe delay characteristics on user experience in short video streaming. Based on analysis in Section~\ref{sec:experimental_results}, we identify four key factors affecting user experience, which are the number of swipe delays, the swipe delay duration, the starting position of each delay event, and the total media duration. Let $N$ and $N^d$ respectively denotes the number of videos and the number of swipe delays in a short video streaming session. The total media duration, denoted $T$, is calculated as the total playback time of all short videos, i.e., $T=\sum_{i=1}^N \tau_i$. The starting position of the swipe delay at video $i$, denoted $t_i$, is measured as the total media duration of all previously viewed videos: $t_i = \sum_{j=1}^i \tau_j$. The proposed QoE model is formulated as follows:
\begin{equation}
    QoE = \alpha + \beta\sum_{i=1}^{N-1}{ d_i \times e^{\lambda t_i/T}} + \gamma N^d,
\end{equation}
where $\alpha$ is a constant representing the baseline QoE without any delays, and $\beta, \lambda,$ and $\gamma$ are the model's coefficients. Inspired by~\cite{exponential1,exponential2}, we model the recency effect using an exponential time-decay weighting of delay durations ($d_i$) as shown in the second term of Eq. (1), where larger weights are assigned to delays that occur closer to the end of the viewing session. The model's parameters are decided to minimize the Mean Squared Error (MSE) between predicted and actual MOS.


\begin{table}[t]
\caption{Coefficients of the Proposed QoE Model}
\begin{center}
\begin{tabular}{|c|c|c|c|c|c|c|}
\hline
\textbf{$\alpha$} & \textbf{$\beta$} & \textbf{$\lambda$} & \textbf{$\gamma$}\\
\hline
4.52 & -0.10 & 0.55 & -0.23\\
\hline
\end{tabular}
\label{tab:model_coefficients}
\end{center}
\end{table}

\begin{table}[t]
\centering
\caption{Model Performance Comparison. Bold numbers indicate the best performance on the test set.}
\resizebox{\linewidth}{!}{
\begin{tabular}{|l|c|c|c|c|c|c|}
\hline
\textbf{Model} & \textbf{Split} & \textbf{Slope} & \textbf{Intercept}& \textbf{RMSE $\downarrow$} & \textbf{PCC $\uparrow$} & \textbf{SROCC $\uparrow$} \\
\hline
Kooij's & Test & 0.748 &  1.830 & 0.589 & 0.759 & 0.834 \\
Ho\ss{}feld's & Test & 1.010 &  -0.170 & 0.752 & 0.532 & 0.651 \\
Tran's & Test & 0.185 &  2.973 & 0.741 & 0.581 & 0.553 \\
CQM & Test & 0.191 &  2.980 & 0.728 & 0.602 & 0.583 \\
P.1203.3 & Test & 1.347 &  -1.066 & 0.715 & 0.595 & 0.537 \\
OLS Cat & Test & 0.632 &  0.712 & 0.845 & 0.333 & 0.329 \\\hline
\multirow{2}{*}{\textbf{Ours}}  & Train & --- & --- &  0.269 & 0.954  & 0.959  \\
  & Test  & --- & --- &  \textbf{0.279} & \textbf{0.953}  & \textbf{0.949}  \\
\hline
\end{tabular}}
\label{tab:modelper}
\vspace{-14pt}
\end{table}

To quantify the performance of the proposed QoE model, we employ three widely used metrics: Root Mean Square Error (RMSE), Pearson Correlation Coefficient (PCC), and Spearman Rank-Order Correlation Coefficient (SROCC)~\cite{ICME_RMSE,ICME_RMSE2,ICME_RMSE3}. RMSE captures the absolute deviation between the predicted scores and the MOSs, while PCC assesses their linear relationship, and SROCC evaluates their ranked correlation. 

The dataset is randomly partitioned into training (80\%) and test (20\%) subsets. The training set is used to learn the model parameters, whereas the test set serves for assessing model performance. This random partitioning procedure is repeated ten times, and the final results are reported as the average across all runs. Table~\ref{tab:model_coefficients} presents the coefficients of the proposed model obtained from the dataset split that achieves the highest PCC on the test set among the ten partitions.


We also evaluate six reference models, namely \textit{Kooij's~\cite{kooij2006perceived}}, \textit{Ho\ss{}feld's}~\cite{Hossfelds}, \textit{Tran's}~\cite{trans}, \textit{CQM}~\cite{CQM},  \textit{P.1203.3}~\cite{p1203_1,p1203_2,p1203_3}, and \textit{OLS Cat}~\cite{CAT}. Notably, Kooij's model was originally developed to measure QoE of zapping time in IPTV, whereas the other models were designed for traditional video streaming rather than short video streaming. Since Kooij's and Ho\ss{}feld's models support only a single stalling or zapping event, we use the total delay duration as their input. 
To ensure a fair comparison, all model parameters are configured according to the values reported in the original publications. To account for potential variations in subjective testing environments and contexts, a first-order linear regression is additionally employed, following~\cite{ITU2012P1401,p1203_3,LSTMTran}.

The resulting regression coefficients (slope and intercept), along with the performance of each model, are summarized in Table~\ref{tab:modelper}. Our model exhibits superior performance, achieving the lowest RMSE (0.279) and the highest PCC (0.953) and SROCC (0.949) on the test set. Among the reference models, Kooij's model, which is designed for the QoE of IPTV applications, achieves the lowest RMSE (0.589) and the highest PCC (0.759) and SROCC (0.834). Meanwhile, the QoE models developed for traditional video streaming show poor performance with $\text{PCC} < 0.7$, $\text{SROCC} < 0.7$, and $\text{RMSE} > 0.7$. These results underscore the necessity for models specifically tailored to short video streaming.


\section{Conclusions}
In this paper, we conduct the first-ever subjective quality evaluation of the effects of swipe delay on users' QoE in short video streaming. Our evaluation consists of 132 swipe delay patterns, which contain one to four swipe delays with durations from 0.25 seconds to 16 seconds. Evaluation results show that the user's QoE is influenced not only by the number of delays and delay duration, but also by the temporal position of the delay within a session. We also introduce the first QoE model specifically designed to predict users' QoE in the presence of swipe delay in short video streaming. The proposed model shows superior performance compared to six traditional QoE models. In future work, we will extend the proposed QoE model to include the effects of other factors, such as video quality, inter-video quality variation, viewing devices, and content characteristics.
\label{sec:conclusion}
\bibliographystyle{IEEEbib}
\bibliography{icme2026references}

@online{threedelaylimit,
  author = {Jakob Nielsen},
  title = {Response Times: The 3 Important Limits},
  year = {1993},
  url = {https://www.nngroup.com/articles/response-times-3-important-limits/},
  urldate = {2025-10-10}
}

@article{WebQoESurvey,
  title={Survey of research on quality of experience modelling for web browsing},
  author={Barakovi{\'c}, Sabina and Skorin-Kapov, Lea},
  journal={Quality and User Experience},
  volume={2},
  number={1},
  pages={6},
  year={2017},
  publisher={Springer}
}

@article{CloudGameQoE,
title = {Gaming in the clouds: QoE and the users’ perspective},
journal = {Mathematical and Computer Modelling},
volume = {57},
number = {11},
pages = {2883-2894},
year = {2013},
note = {Information System Security and Performance Modeling and Simulation for Future Mobile Networks},
issn = {0895-7177},
doi = {https://doi.org/10.1016/j.mcm.2011.12.014},
url = {https://www.sciencedirect.com/science/article/pii/S0895717711007771},
author = {Michael Jarschel and Daniel Schlosser and Sven Scheuring and Tobias Hoßfeld},
}

@inproceedings{Measurement_Study_2025,
author = {Kara, Burak and Simon, Gwendal and Demir, Ayse B. and Begen, Ali C. and Agboma, Florence},
title = {Beyond Swiping through Short-Form Videos},
year = {2025},
booktitle = {Proc. of the 4th Mile-High Video Conf.},
pages = {13–18},
numpages = {6},
address = {Denver, USA},
}

@inproceedings{DeLoad,
author = {Liu, Tong and Fan, Zhiwei and Peng, Guanyan and Zhang, Haodan and Zhang, Yucheng and Wang, Zhen and Xie, Pengjin and Liu, Liang},
title = {DeLoad: Demand-Driven Short-Video Preloading with Scalable Watch-Time Estimation},
year = {2025},
booktitle = {Proc. of the 33rd ACM Int. Conf. on Multimedia},
pages = {6801–6809},
numpages = {9},
address = {Dublin, Ireland}
}

@misc{ITU-T-P910,
  author  = {Int. Telecommunication Union (ITU)},
  title        = {{P.910 : Subjective video quality assessment methods for multimedia applications }},
  year         = {2023}
}

@ARTICLE{QoEdatabase2,
  author={Chen, Chao and Choi, Lark Kwon and de Veciana, Gustavo and Caramanis, Constantine and Heath, Robert W. and Bovik, Alan C.},
  journal={IEEE Trans. on Image Processing}, 
  title={Modeling the Time—Varying Subjective Quality of HTTP Video Streams With Rate Adaptations}, 
  year={2014},
  volume={23},
  number={5},
  pages={2206-2221},
  doi={10.1109/TIP.2014.2312613}}

@article{QoEdatabase1,
  title={Study of temporal effects on subjective video quality of experience},
  author={Bampis, Christos George and Li, Zhi and Moorthy, Anush Krishna and Katsavounidis, Ioannis and Aaron, Anne and Bovik, Alan Conrad},
  journal={IEEE Trans. on Image Processing},
  volume={26},
  number={11},
  pages={5217--5231},
  year={2017},
  publisher={IEEE}
}

@article{QoE2014survey,
  title={A survey on quality of experience of HTTP adaptive streaming},
  author={Seufert, Michael and Egger, Sebastian and Slanina, Martin and Zinner, Thomas and Ho{\ss}feld, Tobias and Tran-Gia, Phuoc},
  journal={IEEE Communications Surveys \& Tutorials},
  volume={17},
  number={1},
  pages={469--492},
  year={2014},
  publisher={IEEE}
}

@article{Tiktok_recommend_alg,
  title={Recommendation algorithm in TikTok: Strengths, dilemmas, and possible directions},
  author={Wang, Pengda},
  journal={Int'l J. Soc. Sci. Stud.},
  volume={10},
  pages={60},
  year={2022},
  publisher={HeinOnline}
}

@inproceedings{SwipeAlong,
author = {Zhu, Shangyue and Karagioules, Theo and Halepovic, Emir and Mohammed, Alamin and Striegel, Aaron D.},
title = {Swipe along: A Measurement Study of Short Video Services},
year = {2022},
booktitle = {Proc. of the 13th ACM Multimedia Systems Conf.},
pages = {123–135},
numpages = {13},
location = {Athlone, Ireland},
series = {MMSys '22}
}

@ARTICLE{Zhang2022,  author={Zhang, Yuming and Liu, Yan and Guo, Lingfeng and Lee, Jack Y. B.},  journal={IEEE Trans. on Mobile Computing},   title={Measurement of a Large-Scale Short-Video Service Over Mobile and Wireless Networks},   year={2022},  volume={},  number={},  pages={1-1},  doi={10.1109/TMC.2021.3139893}}

@inproceedings{kooij2006perceived,
  title={Perceived quality of channel zapping},
  author={Kooij, Robert E and Kamal, Ahmed and Brunnstrom, Kjell},
  booktitle={Communication Systems and Networks},
  pages={156--159},
  year={2006}
}

@inproceedings{nguyen2022,
  title={Network-aware prefetching method for short-form video streaming},
  author={Nguyen, Duc V. and others},
  booktitle={2022 IEEE 24th Int. Workshop on Multimedia Signal Processing (MMSP)},
  pages={1--5},
  year={2022},
  organization={IEEE}
}

@inproceedings{LiveClip,
author = {He, Jianchao and Hu, Miao and Zhou, Yipeng and Wu, Di},
title = {LiveClip: towards intelligent mobile short-form video streaming with deep reinforcement learning},
year = {2020},
isbn = {9781450379458},
booktitle = {Proc. of the 30th ACM Workshop on Network and Operating Systems Support for Digital Audio and Video},
pages = {54–59},
numpages = {6},
address = {Istanbul, Turkey},
series = {NOSSDAV '20}
}

@INPROCEEDINGS{CI_2,
  author={Lyko, Tomasz and Elkhatib, Yehia and Ramdhany, Rajiv and Race, Nicholas},
  booktitle={2023 15th Int. Conf. on Quality of Multimedia Experience }, 
  title={Differential QoE in Picture-in-Picture Gaming Videos: A Subjective Study}, 
  year={2023},
  address={Ghent, Belgium},
  number={},
  pages={221-223},
  doi={10.1109/QoMEX58391.2023.10178572}}

@ARTICLE{SOS_006to018,
  author={Gao, Yixuan and Min, Xiongkuo and Cao, Yuqin and Liu, Xiaohong and Zhai, Guangtao},
  journal={IEEE Trans. on Circuits and Systems for Video Technology}, 
  title={No-Reference Image Quality Assessment: Obtain MOS From Image Quality Score Distribution}, 
  year={2025},
  volume={35},
  number={2},
  pages={1840-1854},
  doi={10.1109/TCSVT.2024.3485684}}

@ARTICLE{exponential2,
  author={Fiedler, Markus and Hossfeld, Tobias and Tran-Gia, Phuoc},
  journal={IEEE Network}, 
  title={A generic quantitative relationship between quality of experience and quality of service}, 
  year={2010},
  volume={24},
  number={2},
  pages={36-41},
  doi={10.1109/MNET.2010.5430142}}

@ARTICLE{exponential1,
author={Pierre LEBRETON, Kazuhisa YAMAGISHI},
journal={IEICE Trans. on Communications},
title={Transferring Adaptive Bit Rate Streaming Quality Models from H.264/HD to H.265/4K-UHD},
year={2019},
volume={E102-B},
number={12},
pages={2226-2242},
doi={10.1587/transcom.2019EBP3045},
ISSN={1745-1345},
month={December}}

@INPROCEEDINGS{SOS_0244,
  author={Ramachandra Rao, Rakesh Rao and Herb, Benjamin and Takala, Helmi-Aurora and Mohamed Ahmed, Mohamed Tarek and Raake, Alexander},
  booktitle={2024 16th Int. Conf. on Quality of Multimedia Experience }, 
  title={{AVT-VQDB-UHD-1-HDR: An Open Video Quality Dataset for Quality Assessment of UHD-1 HDR Videos}}, 
  year={2024},
  volume={},
  address={Karlshamn, Sweden},
  pages={179-185},
  doi={10.1109/QoMEX61742.2024.10598284}}

@INPROCEEDINGS{SOS_03,
  author={Tious, Amar and Vigier, Toinon and Ricordel, Vincent},
  booktitle={2025 17th Int. Conf. on Quality of Multimedia Experience }, 
  title={{Emerging challenges in subjective quality assessment of volumetric videos in eXtended Reality}}, 
  year={2025},
  volume={},
  address={Madrid, Spain},
  pages={1-7},
  doi={10.1109/QoMEX65720.2025.11219888}}

@INPROCEEDINGS{SOS,
  author={Ho\ss{}feld, Tobias and Schatz, Raimund and Egger, Sebastian},
  booktitle={2011 Third Int. Workshop on Quality of Multimedia Experience}, 
  title={SOS: The MOS is not enough!}, 
  year={2011},
  volume={},
  address={Mechelen, Belgium},
  pages={131-136},
  doi={10.1109/QoMEX.2011.6065690}}

@ARTICLE{Gamora,
  author={Hou, Biao and Yang, Song and Li, Fan and Zhu, Liehuang and Jiao, Lei and Chen, Xu and Fu, Xiaoming},
  journal={IEEE Trans. on Parallel and Distributed Systems}, 
  title={Gamora: Learning-Based Buffer-Aware Preloading for Adaptive Short Video Streaming}, 
  year={2024},
  volume={35},
  number={11},
  pages={2132-2146},
  doi={10.1109/TPDS.2024.3456567}}

@inproceedings {dashlet,
author = {Zhuqi Li and Yaxiong Xie and Ravi Netravali and Kyle Jamieson},
title = {Dashlet: Taming Swipe Uncertainty for Robust Short Video Streaming},
booktitle = {20th USENIX Symp. on Networked Systems Design and Implementation (NSDI 23)},
year = {2023},
isbn = {978-1-939133-33-5},
address = {Boston, MA},
pages = {1583--1599},
url = {https://www.usenix.org/Conf./nsdi23/presentation/li-zhuqi},
month = apr
}

@ARTICLE{Phong2023,
  author={Phong, Nguyen Tien and others},
  journal={IEEE Access}, 
  title={Joint Preloading and Bitrate Adaptation for Short Video Streaming}, 
  year={2023},
  volume={11},
  number={},
  pages={121064-121076},
  doi={10.1109/ACCESS.2023.3328956}}

@INPROCEEDINGS{Hossfelds,
  author={Ho\ss{}feld, T. and Egger, S. and Schatz, R. and Fiedler, M. and Masuch, K. and Lorentzen, C.},
  booktitle={2012 Fourth Int. Workshop on Quality of Multimedia Experience}, 
  title={Initial delay vs. interruptions: Between the devil and the deep blue sea}, 
  year={2012},
  volume={},
  number={},
  pages={1-6},
  address={Melbourne, VIC, Australia}}

@article{CQM,
author = {Tran, Huyen T. T. and Ngoc, Nam Pham and Ho\ss{}feld, Tobias and Seufert, Michael and Thang, Truong Cong},
title = {{Cumulative Quality Modeling for HTTP Adaptive Streaming}},
year = {2021},
issue_date = {February 2021},
publisher = {Association for Computing Machinery},
address = {New York, NY, USA},
volume = {17},
number = {1},
issn = {1551-6857},
url = {https://doi.org/10.1145/3423421},
doi = {10.1145/3423421},
journal = {{ACM Trans. on Multimedia Computing, Communications, and Applications}},
month = apr,
articleno = {22},
numpages = {24},
}

@inproceedings{trans,
  title={{A multi-factor QoE model for adaptive streaming over mobile networks}},
  author={Tran, Huyen TT and Ngoc, Nam Pham and Pham, Anh T and Thang, Truong Cong},
  booktitle={{2016 IEEE Globecom Workshops (GC Wkshps)}},
  pages={1--6},
  year={2016},
  address={Washington, DC, USA}
}

@inproceedings{yins,
  title={{A control-theoretic approach for dynamic adaptive video streaming over HTTP}},
  author={Yin, Xiaoqi and Jindal, Abhishek and Sekar, Vyas and Sinopoli, Bruno},
  booktitle={Proc. of the 2015 ACM Conf. on special interest group on data communication},
  pages={325--338},
  year={2015},
  address = {London, United Kingdom},
}

@INPROCEEDINGS{CAT,
  author={Schleicher, Johannes and Wehner, Nikolas and Ho\ss{}feld, Tobias and Seufert, Michael},
  booktitle={2024 16th Int. Conf. on Quality of Multimedia Experience }, 
  title={{(Not) The Sum of Its Parts: Relating Individual Video and Browsing Stimuli to Web Session QoE}}, 
  year={2024},
  pages={104-110},
  address={Karlshamn, Sweden}}

@inproceedings{p1203_1,
 address = {Erfurt},
 author = {Raake, Alexander and Garcia, Marie-Neige and Robitza, Werner and List, Peter and Göring, Steve and Feiten, Bernhard},
 booktitle = {Ninth Int. Conf. on Quality of Multimedia Experience },
 doi = {10.1109/QoMEX.2017.7965631},
 isbn = {978-1-5386-4024-1},
 month = {May},
 publisher = {IEEE},
 title = {{A bitstream-based, scalable video-quality model for HTTP adaptive streaming: ITU-T P.1203.1}},
 year = {2017}
 }

@INPROCEEDINGS{largescale,
  author={Rao, Rakesh Rao Ramachandra and Goring, Steve and Fremerey, Stephan and Keller, Dominik and Raake, Alexander},
  booktitle={2025 17th Int. Conf. on Quality of Multimedia Experience }, 
  title={{A Large-Scale Evaluation of Subject Rating Behaviour in Visual Quality Assessment Studies}}, 
  year={2025},
  volume={},
  address={Madrid, Spain},
  pages={1-7},
  doi={10.1109/QoMEX65720.2025.11219954}}

@INPROCEEDINGS{ICME_RMSE3,
  author={Zhu, Jingwen and Chen, Yixu and Wei, Hai and Sethuraman, Sriram and Wu, Yongjun},
  booktitle={2025 IEEE Int. Conf. on Multimedia and Expo (ICME)}, 
  title={Video Quality Assessment for Resolution Cross-Over in Live Sports}, 
  year={2025},
  volume={},
  number={},
  pages={1-6},
  address={Nantes, France}}

@INPROCEEDINGS{ICME_RMSE2,
  author={Zhu, Xilei and Duan, Huiyu and Yang, Liu and Zhu, Yucheng and Min, Xiongkuo and Zhai, Guangtao and Callet, Patrick Le},
  booktitle={2025 IEEE Int. Conf. on Multimedia and Expo (ICME)}, 
  title={ESVQA: Perceptual Quality Assessment of Egocentric Spatial Videos}, 
  year={2025},
  volume={},
  number={},
  pages={1-6},
  address={Nantes, France}}

@INPROCEEDINGS{ICME_RMSE,
  author={Li, Baojun},
  booktitle={2025 IEEE Int. Conf. on Multimedia and Expo Workshops (ICMEW)}, 
  title={{Multimodal Large Language Model for HDR \& SDR Video Quality Measurement}}, 
  year={2025},
  volume={},
  number={},
  pages={1-3},
  address={Nantes, France}}

@misc{ITU2012P1401,
  author  = {Int. Telecommunication Union (ITU)},
  title        = {{ITU-T Recommendation P.1401: Methods, Metrics and Procedures for Statistical Evaluation, Qualification and Comparison of Objective Quality Prediction Models}},
  year         = {2012}
}

@misc{p1203_3,
  title        = {{ITU-T Recommendation P.1203.3: Models and Tools for Quality Assessment of Streamed Media}},
  author  = {Int. Telecommunication Union (ITU)},
  year         = {2020},
note = {{https://github.com/itu-p1203/itu-p1203/}}
}

@ARTICLE{LSTMTran,
  author={Tran, Huyen T. T. and Nguyen, Duc V. and Ngoc, Nam Pham and Thang, Truong Cong},
  journal={IEEE Trans. on Circuits and Systems for Video Technology}, 
  title={{Overall Quality Prediction for HTTP Adaptive Streaming Using LSTM Network}}, 
  year={2021},
  volume={31},
  number={8},
  pages={3212-3226},
  doi={10.1109/TCSVT.2020.3035824}}

@inproceedings{p1203_2,
 address = {Amsterdam},
 author = {Robitza, Werner and Göring, Steve and Raake, Alexander and Lindegren, David and Heikkilä, Gunnar and Gustafsson, Jörgen and List, Peter and Feiten, Bernhard and Wüstenhagen, Ulf and Garcia, Marie-Neige and Yamagishi, Kazuhisa and Broom, Simon},
 booktitle = {9th ACM Multimedia Systems Conf.},
 doi = {10.1145/3204949.3208124},
 isbn = {9781450351928},
 title = {{HTTP Adaptive Streaming QoE Estimation with ITU-T Rec. P.1203 – Open Databases and Software}},
 year = {2018}
 }


\end{document}